\documentclass[twocolumn,amsmath,amssymb,plb]{revtex4}

\usepackage{epsfig}
\usepackage{hyperref}

\def\al{\alpha}
\def\be{\beta}

\def\de{\delta}

\def\th{\theta}

\def\ta{\tau}

\def\ph{\phi}

\def\ch{\chi}

\def\De{\Delta}

\def\lsim{\mathrel{\rlap{\lower4pt\hbox{\hskip1pt$\sim$}}
    \raise1pt\hbox{$<$}}}
\def\gsim{\mathrel{\rlap{\lower4pt\hbox{\hskip1pt$\sim$}}
    \raise1pt\hbox{$>$}}}

\newcommand{\beq}{\begin{equation}}
\newcommand{\eeq}{\end{equation}}
\newcommand{\bea}{\begin{eqnarray}}
\newcommand{\eea}{\end{eqnarray}}
\newcommand{\bit}{\begin{itemize}}
\newcommand{\eit}{\end{itemize}}

\newcommand{\nn}{\nonumber}
\def\nub{\bar\nu}
\def\M#1#2#3{({\cal M}^{(#1)}_{#2})_{#3}}
\def\Nh{{\hat N}}
\def\Ep{{\hat {\cal E}_+}}
\def\gt{\widetilde g} 
\def\Ht{\widetilde H}

\begin{document}

\title{Search for neutrino-antineutrino oscillations with a reactor experiment}

\author{J.S. D\'iaz$^1$, T. Katori$^2$, J. Spitz$^2$, and J.M. Conrad$^2$ }

\affiliation{
$^1$Physics Department, Indiana University, Bloomington, IN 47405, USA\\
$^2$Massachusetts Institute of Technology, Cambridge, MA 02139, USA
}

\begin{abstract}

The disappearance of reactor antineutrinos in the Double Chooz experiment is used to investigate the possibility of neutrino-antineutrino oscillations arising due to the breakdown of Lorentz invariance. We find no evidence for this phenomenon and set the first limits on 15~coefficients describing neutrino-antineutrino mixing within the framework of the Standard-Model Extension.

\end{abstract}
\maketitle
\section{Introduction}
Einstein's theory of special relativity is based on the assumption of
Lorentz invariance--that physical laws are independent
of the orientation and 
propagation speed of a system.
Despite many careful studies, there is at present no compelling experimental evidence for the
breakdown of Lorentz symmetry~\cite{tables}. However, many candidate
theories of quantum gravity can accommodate the spontaneous breaking
of this spacetime symmetry~\cite{SBS_LV}.   These theories have driven the
development of potential Lorentz violation models~\cite{LVmodels}, as well as
experimental methodologies for direct searches~\cite{KM_SB,DKM}.    In the process, it has been observed that the 
interferometric nature of neutrino oscillations makes them sensitive
probes, uniquely suited to address certain models for Lorentz
violation.

This study makes use of the recent observation of electron antineutrino disappearance in reactor
experiments. 
The Double Chooz experiment first reported an indication for the disappearance of antineutrinos propagating $\sim$1050~m 
from two 4.25~MW reactor cores~\cite{DC2012}. 
The Daya Bay~\cite{DB} and
RENO~\cite{RENO} reactor experiments subsequently observed this disappearance at the discovery level. The results are consistent with oscillations within a standard
three-neutrino mixing model~\cite{Thiago, PDG2012}, where the magnitude of the disappearance is
parameterized by the mixing angle $\th_{13}$. The observation of electron neutrino appearance from muon neutrino beams in long baseline accelerator-based experiments~\cite{t2k_nim,minos_nue} is further validation of the discovery of non-zero $\th_{13}$.    

We explore the possibility that the observed reactor disappearance may have two components: traditional three-neutrino oscillations and neutrino-antineutrino oscillations arising due to deviations from exact Lorentz invariance in the neutrino
sector~\cite{LVnu}.
Minute deviations from exact Lorentz invariance could lead to
violations of the conservation of 
angular momentum, triggering neutrino-antineutrino mixing.
Neutrinos are not readily detectable by the reactor experiments, which nominally search for a coincidence signal characteristic of antineutrino
interactions only.  As a result, neutrino-antineutrino oscillations may be
exhibited as disappearance in the data set. Isolating this additional disappearance contribution 
requires an analysis of the antineutrino candidate event energy spectrum. Currently, only the Double
Chooz data can be used for this purpose, as it is the only reactor experiment that
has published and made available their measured energy spectrum with a full error matrix.

This study complements a past test of Lorentz invariance performed with Double Chooz.
The previous analysis involved the search for a sidereal variation among the antineutrino events. 
Bounds were set on coefficients controlling Lorentz-violating
antineutrino-antineutrino 
oscillations using a reactor experiment for the first time~\cite{DC_LV1}.
The search for neutrino-antineutrino mixing in the present work
constitutes a new test of 
Lorentz symmetry in the context of the Standard-Model Extension (SME)~\cite{SME}.

In its most general form, the disappearance of electron antineutrinos is given by
$P_{\nub_e\to\nub_e} = 1-P_{\nub_e\to\nub_{x'}}-P_{\nub_e\to\nu_x}$, 
where $\nub_{x'}=\nub_\mu,\nub_\ta$ and $\nu_x=\nu_e,\nu_\mu,\nu_\ta$.
Since the SME coefficients that modify antineutrino-antineutrino oscillations (${\nub_e\to\nub_{x'}}$) have already been studied by Double Chooz~\cite{DC_LV1}, we focus only on the coefficients that generate the term $P_{\nub_e\to\nu_x}$. Coefficients that produce Lorentz-violating neutrino-neutrino and antineutrino-antineutrino mixing have also been studied by IceCube~\cite{IceCube_LV}, LSND~\cite{LSND_LV}, MiniBooNE~\cite{MiniBooNE_LV}, and MINOS~\cite{MINOS_LV}. The results are tabulated in Ref.~\cite{tables}.

Incorporating Lorentz violation as a perturbative effect over the dominant mass-driven oscillations leads to neutrino-antineutrino mixing appearing as a second-order effect~\cite{DKM}. The oscillation probability can be written as
\beq
P_{\nub_e\to\nub_e}(E,\text{SME}) = P_{\nub_e\to\nub_e}^{(0)}-P_{\nub_e\to\nu_x}^{(2)}~,
\eeq
where $P_{\nub_e\to\nub_e}^{(0)}\approx1-\sin^22\th_{13}\,\sin^2(1.267\De m^2_\text{atm}L/E)$ is the conventional disappearance probability, parameterized by a mixing angle $\theta_{13}$, the atmospheric mass splitting in eV$^2$ ($\De m^2_\text{atm}$), the distance the antineutrino travels in meters ($L$), and its energy in MeV ($E$)~\cite{PDG2012}. 
This approximation is valid for our analysis because the antineutrinos in Double Chooz travel a distance that is too short to be significantly affected by oscillations driven by the solar mass-squared difference; therefore, we neglect the effects of $\De m^2_\odot$.
In the SME, the second-order correction is given by~\cite{DKM}
\beq\label{P(2)}
P_{\nub_e\to\nu_x}^{(2)}=
L^2\bigg|\sum_{c=e,\mu\ta}\sum_{\bar d=\bar e,\bar\mu\bar\ta}
\M{1}{x\bar e}{c\bar d}\,\,\de h_{c\bar d}\bigg|^2~,
\eeq
where the Hamiltonian $\de h$ encodes the coefficients controlling Lorentz violation and the factors $\M{1}{x\bar e}{c\bar d}$ depend on experimental parameters including location, orientation, baseline, and antineutrino energy. These factors also depend on the conventional oscillation parameters~\cite{PDG2012}.
There are 81 different SME coefficients that can lead to independent effects in the oscillation probability, Eq.~\eqref{P(2)}.
In a recent study using data from the MINOS experiment, the 66 coefficients that induce sidereal variations of the oscillation probability have been constrained~\cite{RebelMufson}.
For this reason, we can remove these coefficients from our analysis and study the remaining 15~coefficients whose time-independent effects have not been explored to date. 
The component of the Hamiltonian that remains unconstrained has the explicit form
\beq\label{dh}
\de h_{c\bar d} = 
-i\sqrt2
\Ep^Z \Ht^Z_{c\bar d}
+
i\sqrt2
\Ep^Z \Big(
\gt^{ZT}_{c\bar d}
-\Nh^Z \, \gt^{ZZ}_{c\bar d} 
\Big)E~,
\eeq
where $\Nh^Z$ and $\Ep^Z$ denote the directional factors and neutrino polarization vector, respectively, along the $Z$ axis of the Sun-centered celestial equatorial frame~\cite{SunFrame}, widely used to report results of searches for Lorentz violation. 
These constant factors can be written in terms of the orientation of the neutrino beam ($\th,\ph$) and the colatitude $\ch$ of the experiment~\cite{DKM}.
The Hamiltonian in Eq.~\eqref{dh} also includes three coefficients $\Ht^Z_{c\bar d}$ ($c\bar d=e\bar\mu, e\bar\ta, \mu\bar\ta$) that control Lorentz violation while preserving CPT invariance, and twelve coefficients $\gt^{\al\be}_{c\bar d}$ ($c\bar d=e\bar e, \mu\bar\mu,\ta\bar\ta, e\bar\mu, e\bar\ta, \mu\bar\ta$ and $\al\be=ZT, ZZ$) directing both Lorentz and CPT violation~\cite{CPTv}. These coefficients are complex numbers and the form of the probability \eqref{P(2)} shows that our analysis is sensitive to their absolute values.

The Hamiltonian in Eq.~\eqref{dh} involves an unconventional energy dependence. Contrary to the ordinary mass-driven oscillations controlled by a Hamiltonian that depends inversely on the neutrino energy, Eq.~\eqref{dh} shows that CPT-preserving Lorentz violation introduces energy-independent oscillations and CPT-violating Lorentz violation leads to oscillations that grow linearly with $E$. For this reason, a fit to the energy spectrum allows the potential separation of the three types of contributions to the Hamiltonian due to their characteristic energy dependence.

\section{Analysis}
This analysis is based on a data release by the Double Chooz collaboration, coinciding with their publication describing evidence for non-zero $\th_{13}$~\cite{data_release, DC2012}. The analysis uses 8249 electron antineutrino candidate events detected over about one year with Double Chooz's liquid scintillator based far detector. The antineutrinos are detected with the inverse beta decay reaction $\overline{\nu}_e p \rightarrow e^+ n$, which creates a coincidence of signals separated in time from the initial positron interaction followed by a delayed neutron capture on a gadolinium or hydrogen nucleus~\cite{DC_hydrogen}. 
The dominant backgrounds are spallation products ($^9$Li and $^8$He), stopping muons, and cosmic- and radioactivity-induced fast neutrons. However, these backgrounds are constrained with an $in-situ$ measurement using reactor-off data~\cite{DC_off}. A complete description of the Double Chooz experiment and data analysis can be found in Ref.~\cite{DC2012}.

The data release provides the collaboration's predictions for non-oscillated signal and background energy spectra, error matrices, and data associated with the measurement periods employed for their nominal $\th_{13}$ analysis. The covariance matrices take into account correlated and uncorrelated uncertainties associated with the detector response, background prediction, statistics, and knowledge of the reactor flux. The data release information is used here in order to search for neutrino-antineutrino oscillations and the breakdown of Lorentz invariance. Double Chooz breaks up the data into two ``integration periods". One period utilizes data taken with both reactors on and one period utilizes data taken with one reactor at $<$20\% thermal power. The antineutrino spectral information then comes in terms of the prompt positron's visible energy from 0.7-12.2~MeV. This is converted to antineutrino energy with $E\cong E_{\mathrm{prompt}}+0.78~\mathrm{MeV}$. Note that we employ Double Chooz's best fit central values and uncertainties, rather than the before-fit predictions, from their analysis for the backgrounds, energy scale, and atmospheric mass splitting. This is consistent with the data release, prediction, and knowledge of the reactor flux. 

Before employing the data release in searching for Lorentz violation, we successfully reproduced Double Chooz's $\theta_{13}$ result. With this confirmation, we proceed to extract the Lorentz violating coefficients describing neutrino-antineutrino oscillations. From the structure of Eq.~\eqref{dh} we can write $\de h_{c\bar d}=\text{C}^{(0)}_{c\bar d}+\text{C}^{(1)}_{c\bar d}E$ in order to fit the three factors $\text{C}^{(0)}_{c\bar d}$ that have units of energy and the six dimensionless factors $\text{C}^{(1)}_{c\bar d}$. It is important 
to notice that the absence of a significant signal of Lorentz violation could 
appear due to a rather unlikely cancellation between different coefficients. 
Nonetheless, each of the components of the effective Hamiltonian appears 
coupled to a different factor $(\mathcal{M}^{(1)}_{x\bar e})_{c\bar d}$, whose convoluted energy dependence makes any cancellation possible only at a 
given neutrino energy. For this reason, the absence of a positive signal in the entire energy spectrum used in this study allows us to conclude that each component is individually small; therefore, we can set upper limits on the factors $\text{C}^{(0)}_{c\bar d}$ and $\text{C}^{(1)}_{c\bar d}$ by considering only one component of the Hamiltonian at a time. As an illustration, the contribution to the probability in Eq.~\eqref{P(2)} introduced by the component $\de h_{e\bar e}$ takes the explicit form
\bea
P_{\nub_e\to\nu_x}^{(2)}&=&
L^2E^2\,|\text{C}^{(1)}_{e\bar e}|^2
\Big(
\big|\M{1}{e\bar e}{e\bar e}\big|^2
\nn\\
&&\quad\quad
+
\big|\M{1}{\mu\bar e}{e\bar e}\big|^2
+
\big|\M{1}{\tau\bar e}{e\bar e}\big|^2
\Big)~,
\eea
where the three terms correspond to the oscillation channels $\nub_e\to\nu_e$, $\nub_e\to\nu_\mu$, and $\nub_e\to\nu_\ta$, respectively.
Notice that only the complex factor $\text{C}^{(1)}_{e\bar e}$ appears because the factor $\text{C}^{(0)}_{e\bar e}$ vanishes due to the antisymmetry of the SME coefficient $\Ht^\al_{e\bar e}$ in mixed-flavor space. Similar results appear for the oscillations caused by $\de h_{\mu\bar\mu}$ and $\de h_{\ta\bar\ta}$, 
whereas the off-diagonal components of the Hamiltonian in Eq.~\eqref{dh} produce modifications involving both $\text{C}^{(0)}_{c\bar d}$ and $\text{C}^{(1)}_{c\bar d}$. Due to the symmetry of the oscillation probability equation, the $\mu\bar\mu$ and $\tau\bar\tau$ as well as the $e\bar\mu$ and $e\bar\tau$ fit functions and results are the same. 

\begin{figure}[h]
\begin{centering}
\includegraphics[height=2.3in]{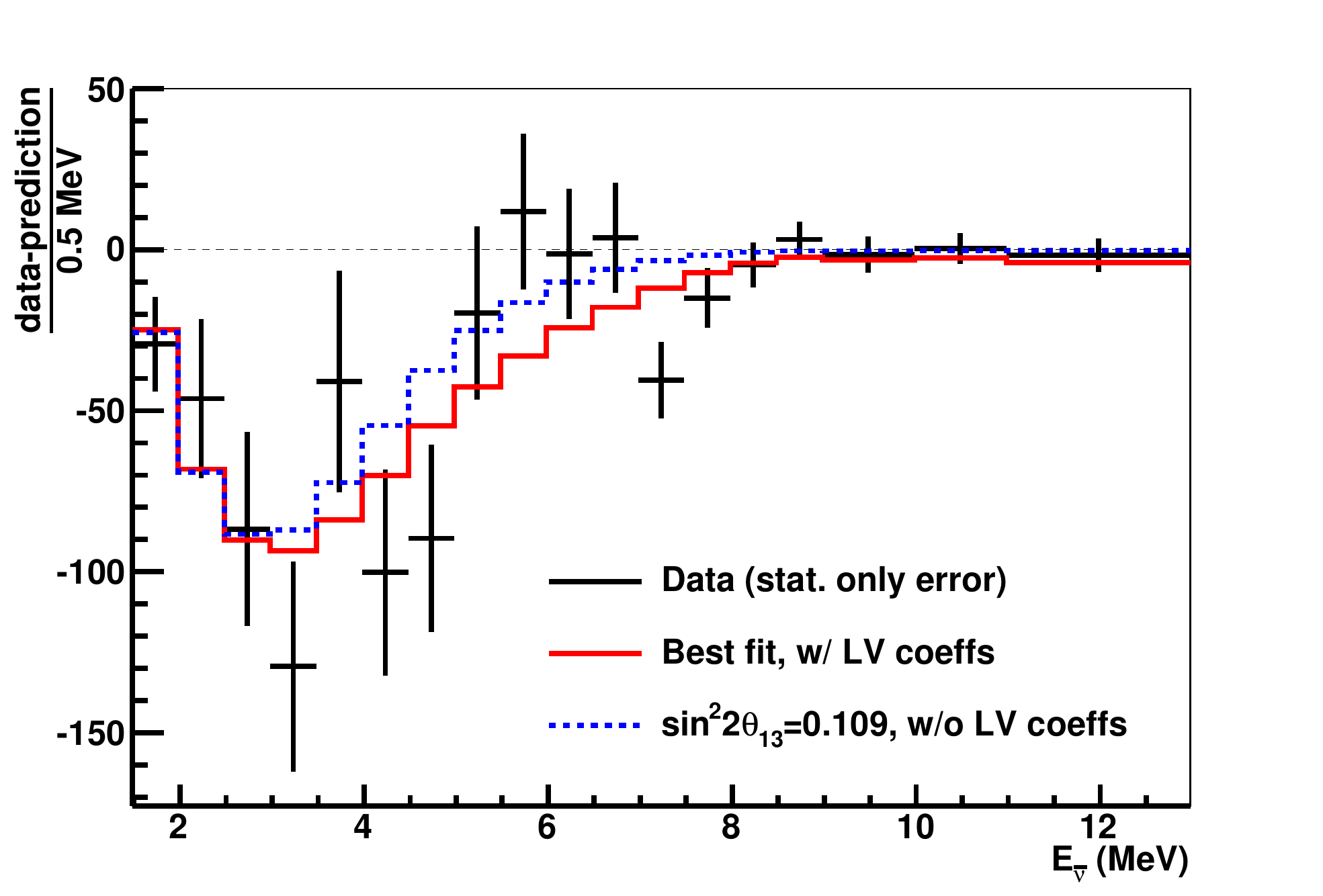} 
\caption{The observed antineutrino rate compared to the prediction as a function of antineutrino energy. The $e\bar e$ best fit results along with the data points and Double Chooz's $\theta_{13}$-only best fit are shown.}
\label{1}
\end{centering}
\end{figure} 

In order to study the three-neutrino oscillation contributions to the
reactor signal it is advantageous to account for current constraints on $\th_{13}$. Since we are testing the possibility of Lorentz violation modifying the antineutrino disappearance probability in Double Chooz, it would be incongruous to use the value of this mixing angle measured through the disappearance channel in reactor experiments. We therefore take T2K's result $\sin^2(2\theta_{13})_{\mathrm{T2K}}=0.088^{+0.049}_{-0.039}$, given for a normal hierarchy, $\delta_{\mathrm{CP}}=0$, $\Delta m^2_{\mathrm{atm}}$=2.4$\times10^{-3}$~eV$^2$, and $\theta_{23}=45^\circ$, and use it as a constraint in the fits~\cite{t2k_nim}. It is likely that the T2K and MINOS appearance measurements are dominated by mass-based ($\th_{13}$) oscillations, rather than Lorentz violation, because they are mutually consistent and yet have very different sensitivities to Lorentz violation themselves given their differing baselines (295~km and 735~km) and neutrino energy spectra ($<$0.6~GeV$>$ and $<$3~GeV$>$).  We do note, however, that the Lorentz-violating oscillation of muon neutrinos into electron antineutrinos could mimic the electron-like appearance signal observed in T2K. We considered this possibility and found that the spectral distribution of the few electron-like events in T2K disfavors this type of Lorentz-violating oscillation because the effects of the coefficients $\Ht^\al_{c\bar d}$ and $\gt^{\al\be}_{c\bar d}$ grow with $L^2$ and $L^2E^2$, respectively. This means that a non-zero value of the SME coefficients would have to be of order $10^{-23}$ or less for the result to be compatible with T2K data. The vastly different baselines and energies of the antineutrinos in reactor experiments make them largely insensitive to possible violations of Lorentz invariance affecting long-baseline experiments. These reasons allow us to take the T2K measurement of $\th_{13}$ to be free of Lorentz-violating effects within the reach of Double Chooz.

We employ a least squares fitting technique for comparing the Monte Carlo signal prediction plus background expectation and the data and extracting the best fit (BF) parameters associated with oscillations. 
The least squares estimator is defined as
\begin{eqnarray}
X^2 =&\sum_{ij}^{N}&
[r_{i,data}-P_{\nub_e \to \nub_e}(E,\text{SME})\cdot r_{i,MC}]
\cdot M_{ij}^{-1}   \nonumber \\
&&\quad\quad \cdot[r_{j,data}-P_{\nub_e \to \nub_e}(E,\text{SME})\cdot r_{j,MC}] \nonumber \\
&&\quad\quad +\frac{[\sin^2(2\theta_{13})-\sin^2(2\theta_{13})_{\mathrm{T2K}}]^2}{\sigma^2_{\mathrm{T2K}}}~,
\end{eqnarray}
\noindent where $i$/$j$ is the bin number (1-36 inclusive), $r_{i,data}$ and $r_{i,MC}$ 
are the data and MC expectation event count vectors, 
$P_{\nub_e \to \nub_e}(E,\text{SME})$ is the energy-dependent oscillation probability 
based on the Lorentz-violating model being considered, 
and $M_{ij}^{-1}$ is the inverse of the total error matrix. 
A pull term constraining the value of 
$\theta_{13}$, based on T2K's result, is introduced as mentioned above.

\begin{figure}[tb]
\begin{centering}
\includegraphics[height=2.3in]{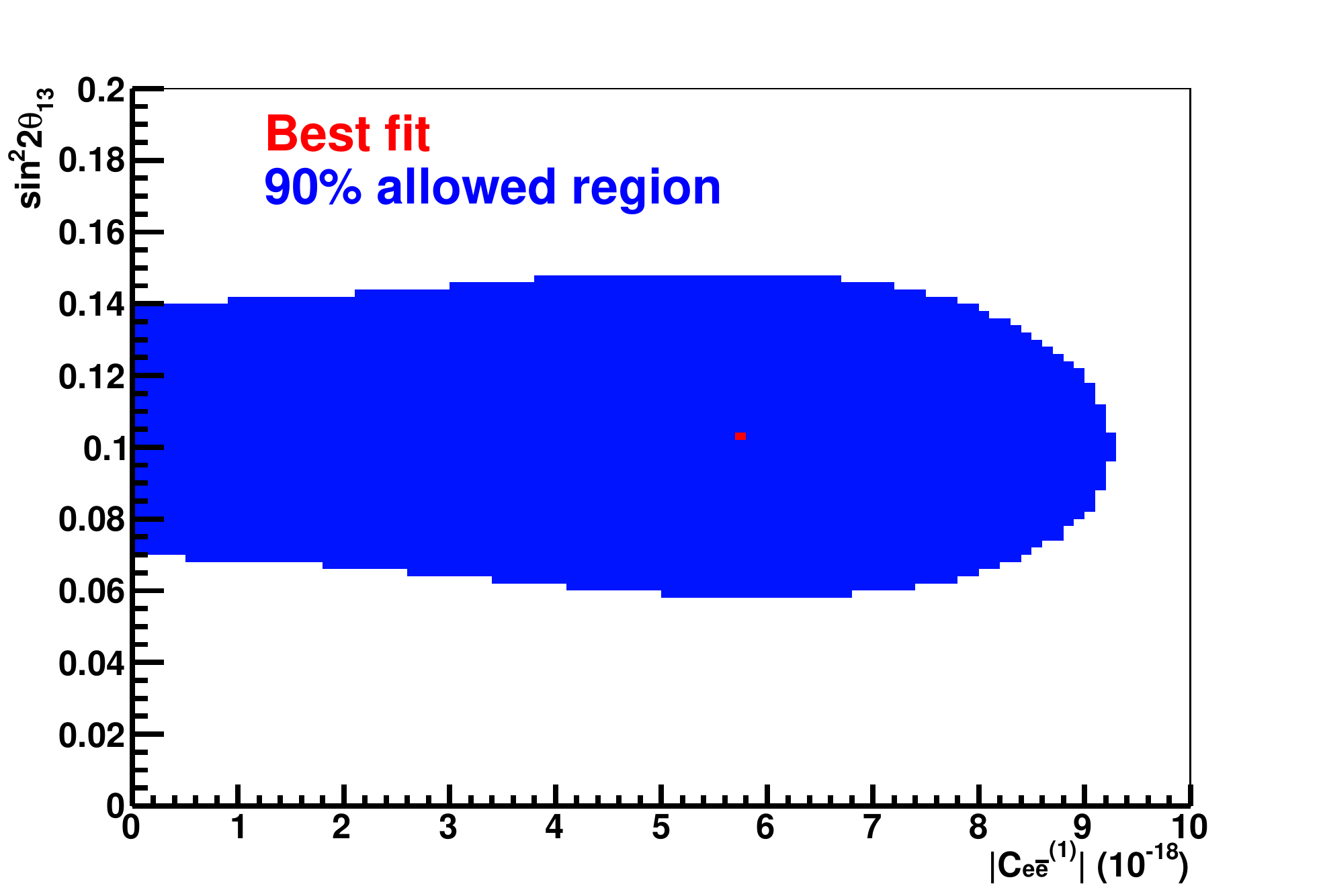} 
\\
\includegraphics[height=2.3in]{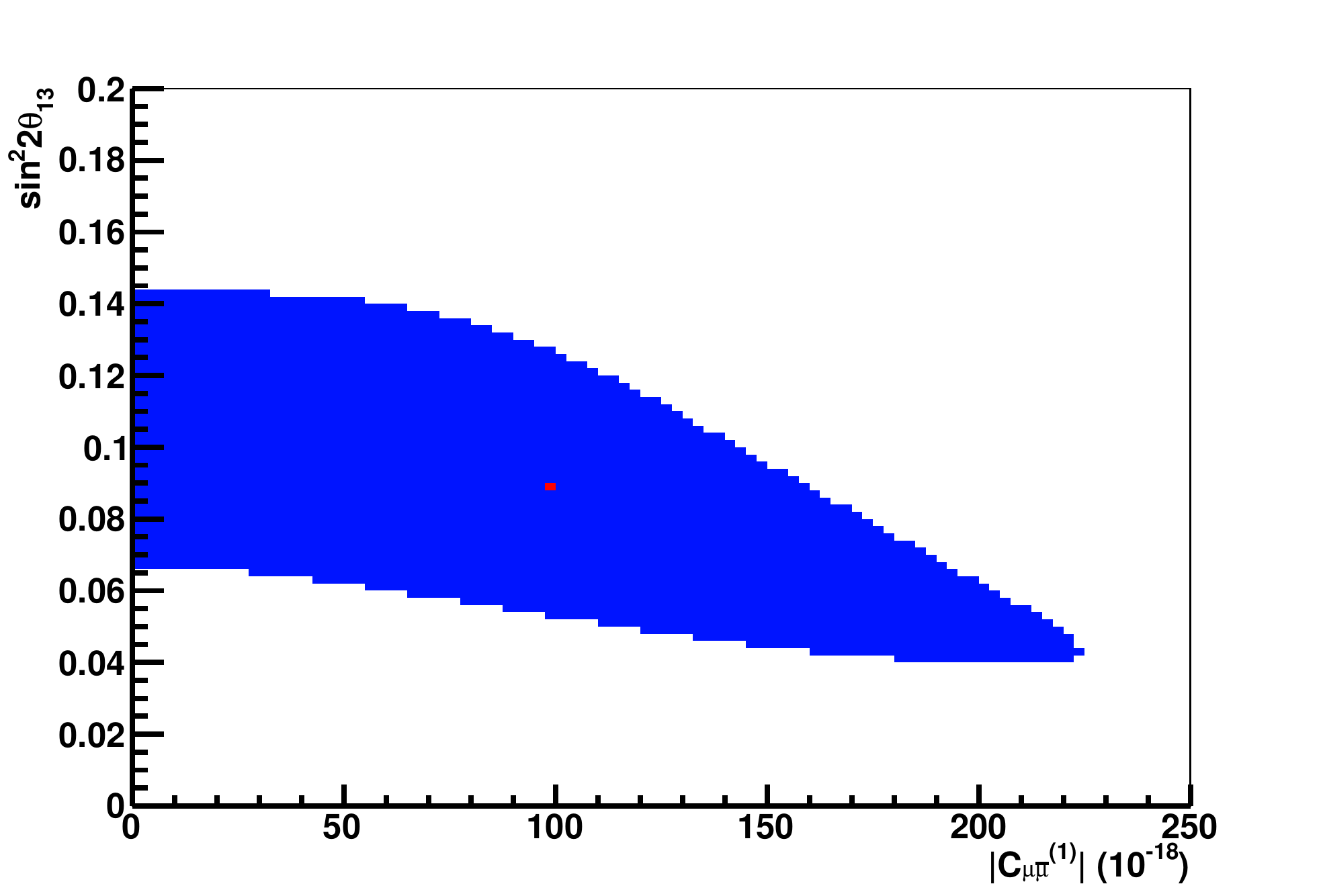} 
\caption{The best fit points and 90\% CL regions for the $e\bar e$ (top) and $\mu\bar\mu$ (or $\tau\bar\tau$, bottom) fits. The $e\bar e$ ($\mu\bar\mu$ or $\tau\bar\tau$) best fit function has $\chi^2/ndf$=42.4/35 (42.6/35).}
\label{2}
\end{centering}
\end{figure}

\begin{figure}[tb]
\begin{centering}
\includegraphics[height=2.3in]{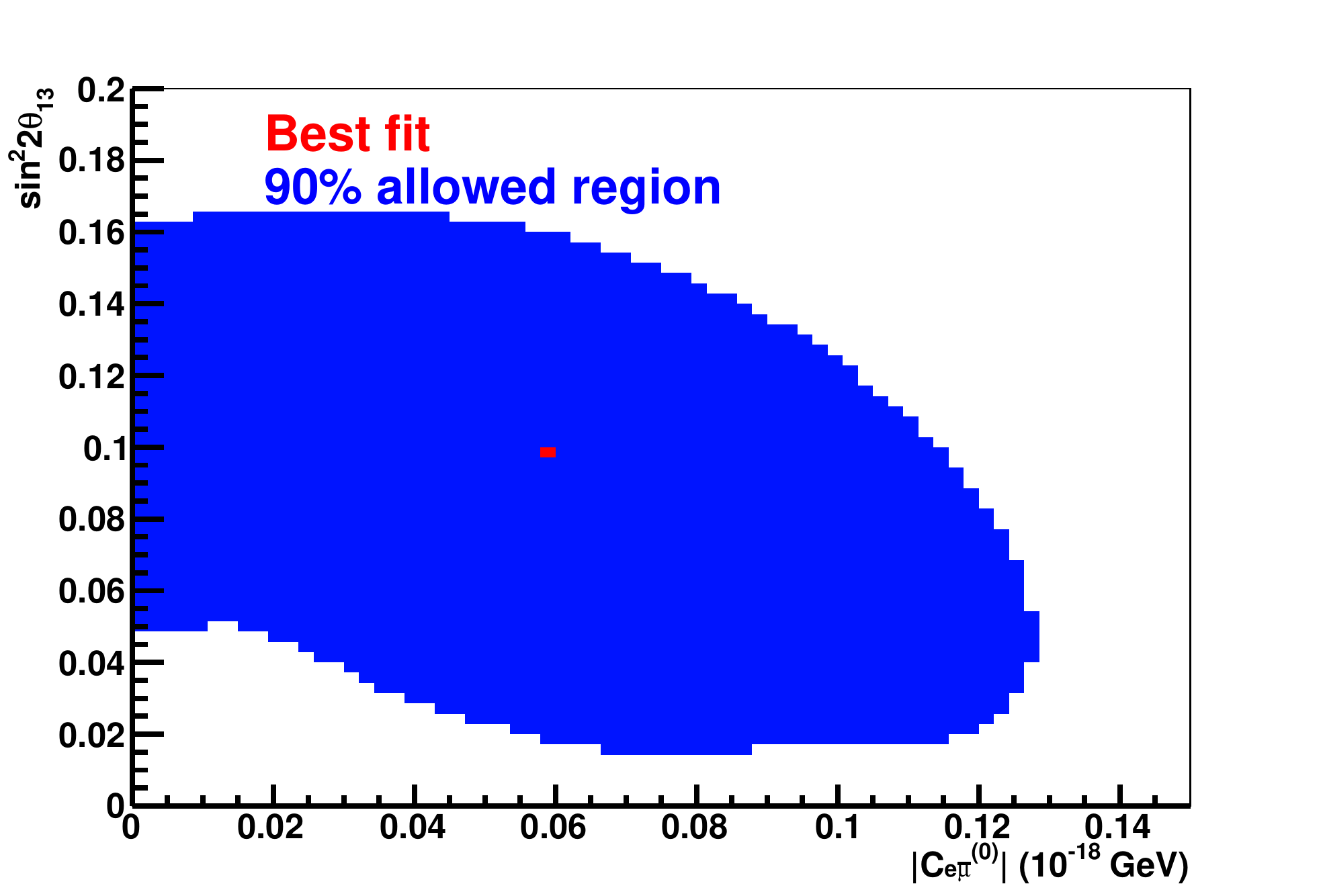} 
\\
\includegraphics[height=2.3in]{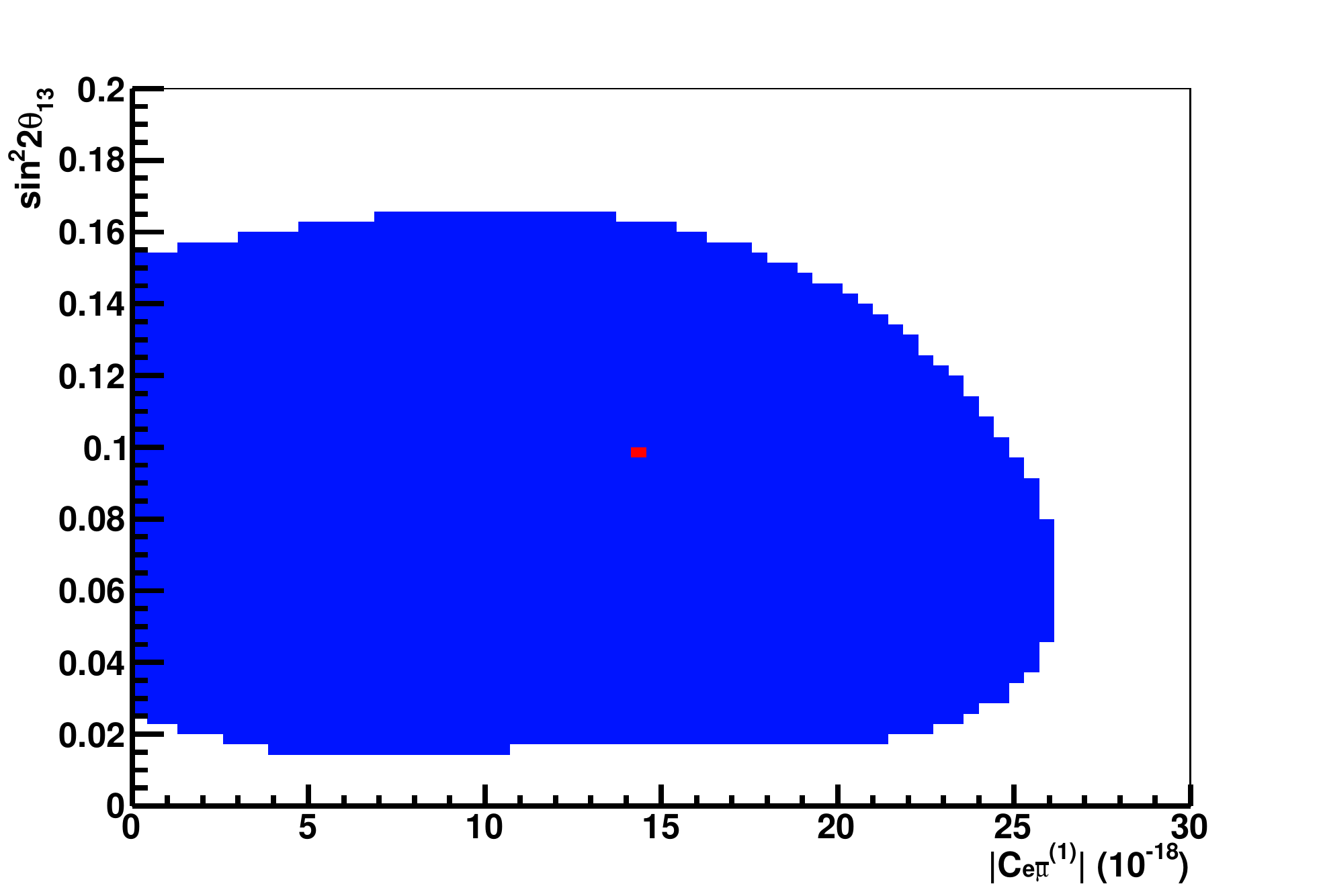} 
\\
\includegraphics[height=2.3in]{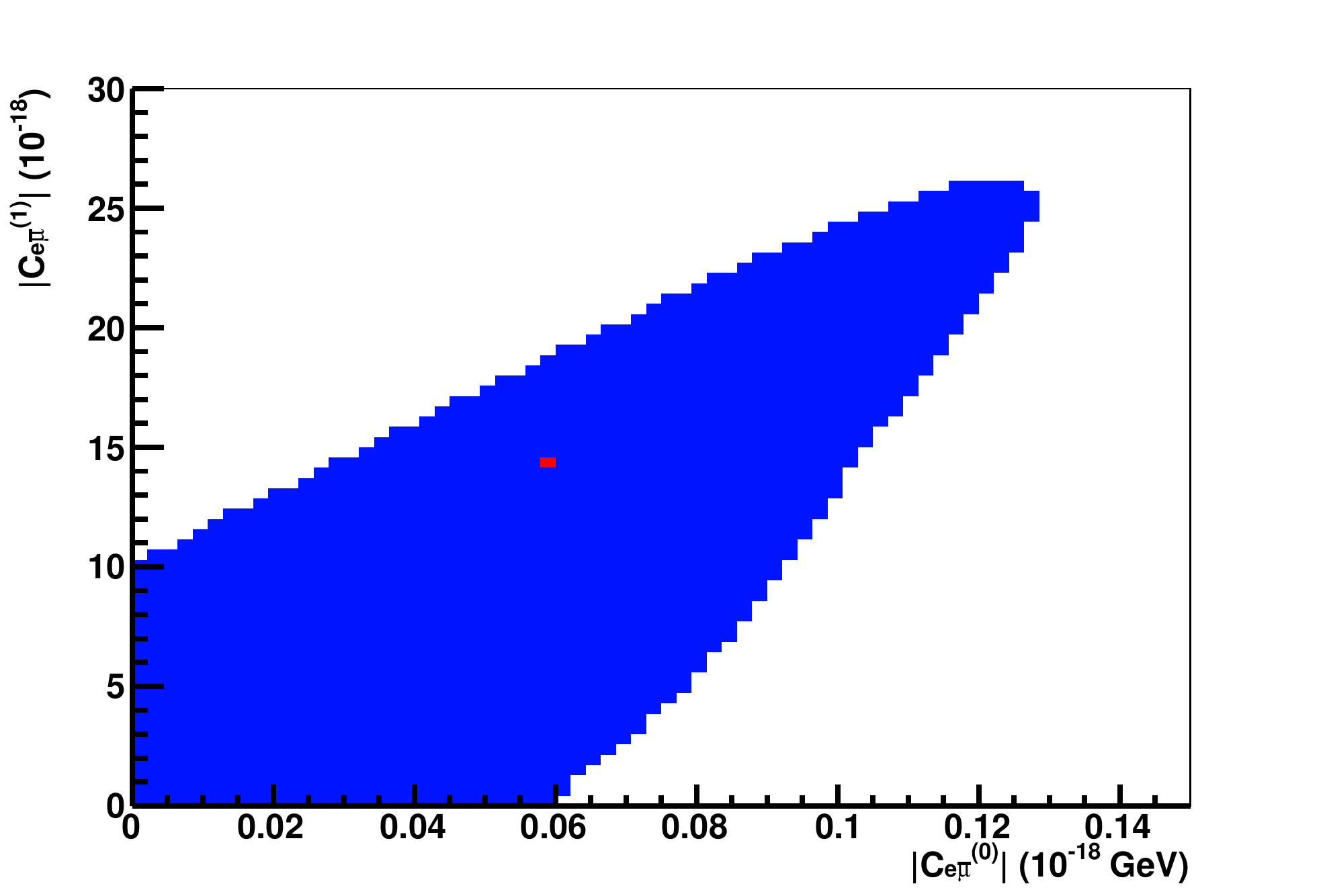} 
\caption{The best fit points and 90\% CL regions for the $e\bar\mu$ (or $e\bar\tau$) fits. The best fit function has $\chi^2/ndf$=42.1/34.}
\label{3}
\end{centering}
\end{figure}

\begin{figure}[tb]
\begin{centering}
\includegraphics[height=2.3in]{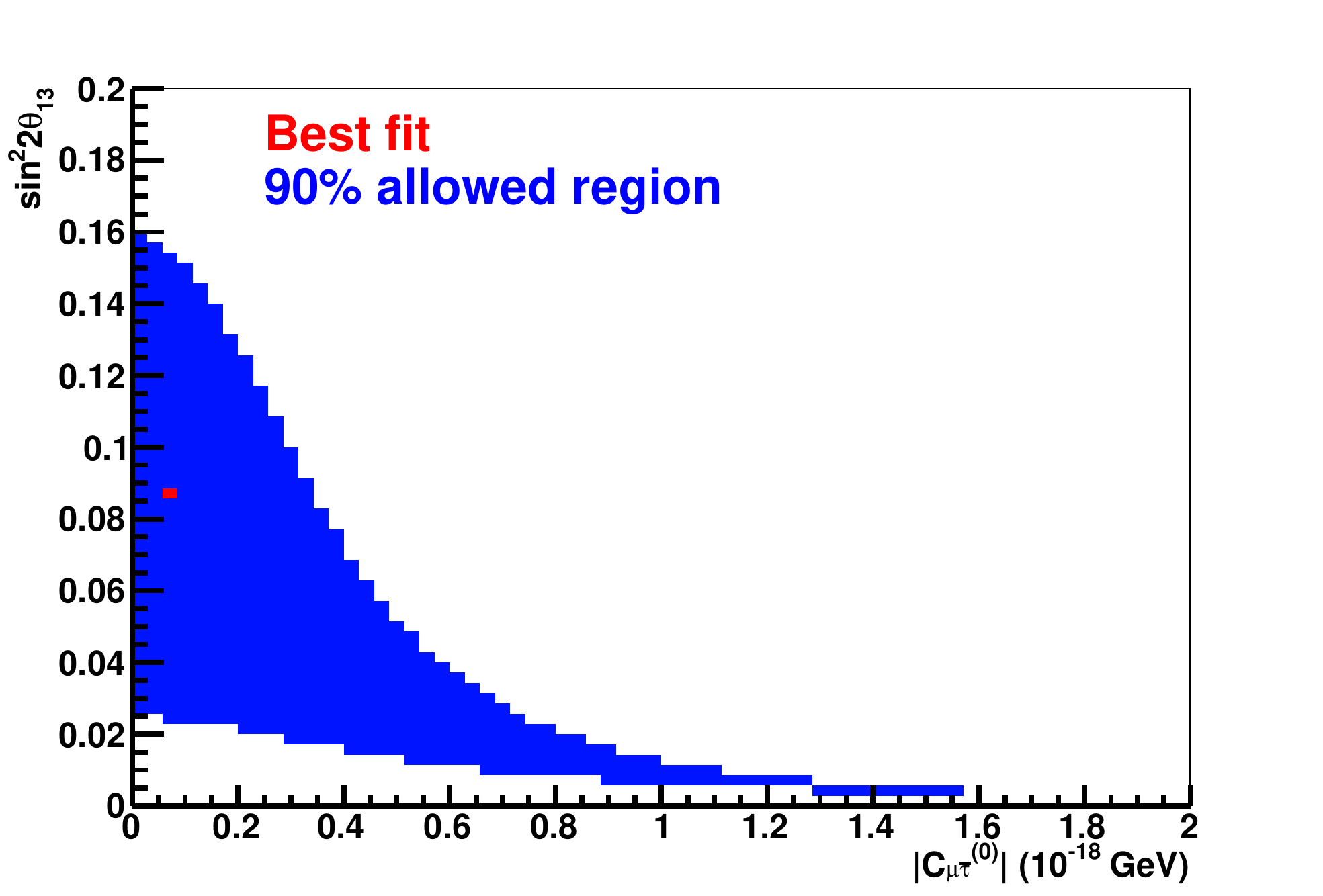} 
\\
\includegraphics[height=2.3in]{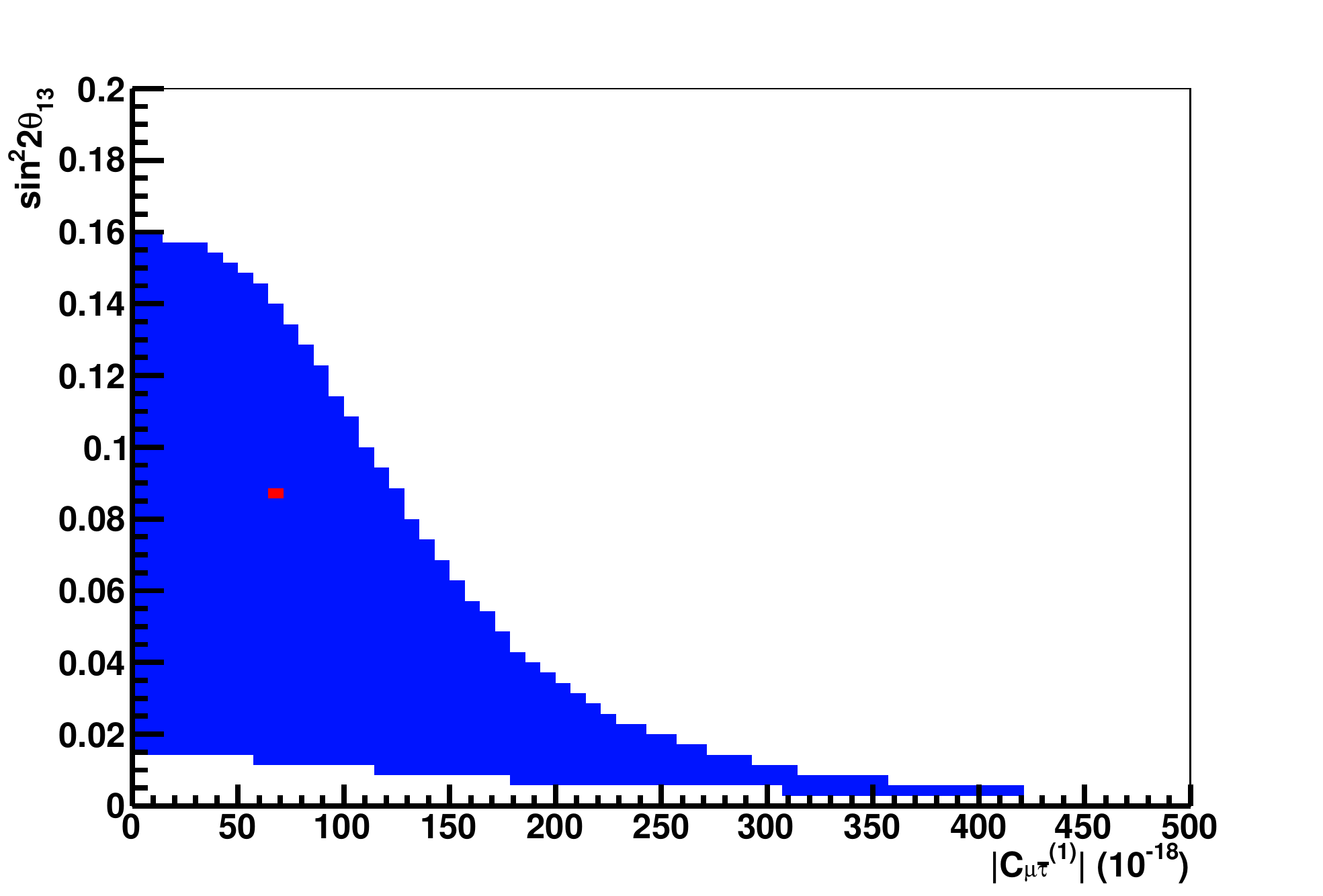} 
\\
\includegraphics[height=2.3in]{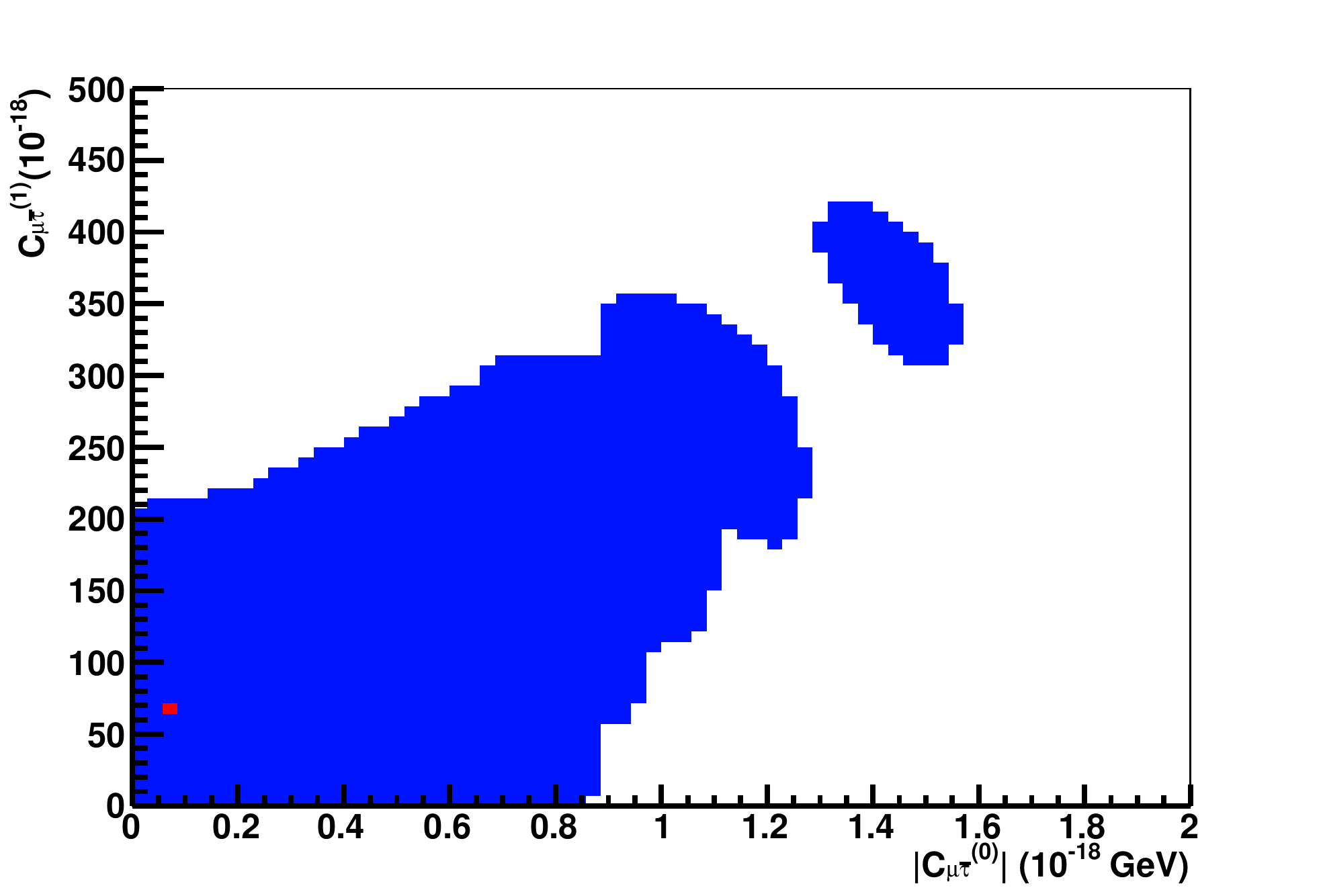} 
\caption{The best fit points and 90\% CL regions for the $\mu\bar\tau$  fits. The best fit function has $\chi^2/ndf$=42.6/34.}
\label{4}
\end{centering}
\end{figure}

The least squares estimator is minimized with the MINUIT software~\cite{MINUIT} to find the best fit point with the relevant set of SME coefficients. The best fit $e\bar e$ results overlaid with the data are shown in Figure~\ref{1} as an example. Although both integration periods (18~bins in energy each) have been simultaneously considered when performing the fits, we report our fit results on one combined-period plot for simplicity. The minimization is checked and the allowed regions around the best fit values are formed with a raster scan technique involving the creation of an $X^2$ map; $X^2$ is determined for each possible combination of parameters. We assume the minimum of $X^2$ follows a $\chi^2$ distribution with $ndf = 36-P $, where $P$ is the number of fit parameters. This assumption is checked using a frequentist study, as described below. Confidence regions are drawn based on a $\Delta \chi^2$, defined as the $X^2$ value at each point minus the $X^2$ at the best fit point. For the one parameter fits [$e\bar e$ and $\mu\bar\mu$ (or $\tau\bar\tau$)], the 90\% confidence region encloses the values which satisfy the condition $\Delta \chi^2<2.71$. The two parameter [$e\bar\mu$ (or $e\bar\tau$) and $\mu\bar\tau$] 90\% confidence regions enclose the values which satisfy the condition $\Delta \chi^2<4.61$. The best fit values and confidence regions for each of the fits are shown in Figures~\ref{2},~\ref{3}, and~\ref{4}.

The confidence regions are verified with the generation of numerous pseudoexperiments drawn based on the Monte Carlo expectation. The distributions are created with the use of the full covariance matrix and represent the statistical and systematic fluctuations expected in data. The pseudoexperiments are generated without any oscillations. Post-generation, the spectral distributions corresponding to each pseudoexperiment are convoluted with the best fit function from the relevant fit to data. After modifying the covariance matrix to account for the statistical error in each pseudoexperiment, the estimator is minimized and the best fit function [$P_{\nub_e\to\nub_e}(E,\text{SME})$] is found for each pseudoexperiment sample. The best fit values for the thousands of simulated pseudoexperiments are then recorded on an $X^2$ map for comparison with the confidence regions previously formed. The fraction of pseudoexperiment best fit points inside of the allowed regions is found to be consistent with 90\% for each of the fits and the regions are substantiated. 

Given that the 90\% CL allowed regions generated based on Double Chooz's data encompass the null, no neutrino-antineutrino oscillation hypothesis in all cases considered, we conclude that there is no evidence for this process and proceed to limit the relevant SME coefficients. The results are shown in Table~\ref{tab:SME1}. Note that since the fit functions are squares of the fit parameters, the best fit points are always duplicated and the sign reversed values are equally reasonable. 

\begin{table}[htb] \begin{center}
%\footnotesize
\begin{tabular}{c|c|c|c|c}
\hline
\hline
\phantom{aaaaa}&\multicolumn{2}{c}{~~~~~~Best fit~~~~~~}
&\multicolumn{2}{|c}{~Upper limit (90\% CL)~}\\\hline
$c\bar d$
& $|\text{C}_{c\bar d}^{(0)}|\,(\text{GeV})$ 
& ~~~~~$|\text{C}_{c\bar d}^{(1)}|$~~~~~
& $|\text{C}_{c\bar d}^{(0)}|\,(\text{GeV})$ 
& ~~~~~$|\text{C}_{c\bar d}^{(1)}|$~~~~~\\ \hline
$e\bar e$   &   $-$ &   5.7 &   $-$ &   9.3   \\
$\mu\bar\mu$    &   $-$ &   100 &   $-$ &   225  \\
$\ta\bar\ta$    &   $-$ &   100 &   $-$ &   225  \\
$e\bar\mu$  &   0.06    &   14  &   0.13    &   26   \\
$e\bar\ta$  &   0.06    &   14  &   0.13    &   26   \\
$\mu\bar\ta$    &   0.07 &   70  &   1.6   &   420  \\
\hline
\hline
\end{tabular}
\end{center}
\caption{Best fit values and upper limits (90\% CL) of the nine factors producing neutrino-antineutrino oscillations. The values are in units of $10^{-18}$.}
\label{tab:SME1}
\end{table}
\begin{table}[htb] \begin{center}
%\footnotesize
\begin{tabular}{c|c|c}
\hline
\hline
    $-  $&  $|\gt^{ZT}_{e\bar e}|   < 9.7\times10^{-18} $&  $|\gt^{ZZ}_{e\bar e}|   < 3.3\times10^{-17} $ \\
    $-  $&  $|\gt^{ZT}_{\mu\bar\mu}|    < 2.3\times10^{-16} $&  $|\gt^{ZZ}_{\mu\bar\mu}|    < 8.1\times10^{-16} $ \\
    $-  $&  $|\gt^{ZT}_{\ta\bar\ta}|    < 2.3\times10^{-16} $&  $|\gt^{ZZ}_{\ta\bar\ta}|    < 8.1\times10^{-16} $ \\
$|\Ht^Z_{e\bar\mu}| <   1.4\times10^{-19}   $&  $|\gt^{ZT}_{e\bar\mu}|  < 2.7\times10^{-17} $&  $|\gt^{ZZ}_{e\bar\mu}|  < 9.3\times10^{-17} $ \\
$|\Ht^Z_{e\bar\ta}| <   1.4\times10^{-19}   $&  $|\gt^{ZT}_{e\bar\ta}|  < 2.7\times10^{-17} $&  $|\gt^{ZZ}_{e\bar\ta}|  < 9.3\times10^{-17} $ \\
$|\Ht^Z_{\mu\bar\ta}| < 1.7\times10^{-18}   $&  $|\gt^{ZT}_{\mu\bar\ta}|    < 4.4\times10^{-16} $&  $|\gt^{ZZ}_{\mu\bar\ta}|    < 1.5\times10^{-15} $ \\
\hline
\hline
\end{tabular}
\end{center}
\caption{Limits for the 15 independent SME coefficients that produce neutrino-antineutrino oscillations. The coefficients for CPT-conserving Lorentz violation $\Ht^Z_{c\bar d}$ are given in units of GeV and the coefficients for CPT-violating Lorentz violation $\gt^{\alpha \beta}_{c\bar d}$ are dimensionless.}
\label{tab:SME2}
\end{table}

Using the location of Double Chooz as well as the orientation of the detector with respect to the reactors, the factors $\text{C}_{c\bar d}^{(0)}$ and $\text{C}_{c\bar d}^{(1)}$ can be written in terms of the SME coefficients with the form
\bea
|\text{C}_{c\bar d}^{(0)}| &=& 0.96\,|\Ht^Z_{c\bar d}|~,
\nn\\
|\text{C}_{c\bar d}^{(1)}| &=&
0.96\,|\gt^{ZT}_{c\bar d}+0.29\gt^{ZZ}_{c\bar d}|~,
\eea
which can be used to set limits on the three individual coefficients for CPT-even Lorentz violation $H^Z_{c\bar d}$. The coefficients for CPT-odd Lorentz violation $g^{TZ}_{c\bar d}$ and $g^{ZZ}_{c\bar d}$ appear in pairs. Even though the fits provide upper limits on combinations of these coefficients, presented in the last column of Table~\ref{tab:SME1}, individual limits on the coefficients are reported after considering each one at a time in these six combinations. We present these individual limits for completeness in Table~\ref{tab:SME2}. The values reported correspond to limits on the absolute value of each coefficient; nonetheless, these limits can also be interpreted as bounds on the modulus of the real and imaginary parts of the corresponding coefficient.

\section{Conclusion}
The reactor-based antineutrino experiments' recent measurement of non-zero $\theta_{13}$ is an important milestone in particle physics and represents the satisfaction of the main prerequisite for a precise determination of the CP-violating phase in the lepton sector. 
These experiments are sensitive to more than just $\theta_{13}$, however. The collections of electron antineutrino events can also be used as a sensitive probe of physics beyond the Standard Model. 
In the present work, we have taken the Double Chooz results, made explicit and clear in the form of a detailed data release, and conducted a search for neutrino-antineutrino oscillations. 
No evidence for this exotic process has been found and we set limits on 15 previously unexplored SME coefficients. This analysis shows that antineutrinos in the Double Chooz experiment are sensitive 
to Lorentz-violating effects suppressed by the Planck-scale ($M_P\simeq10^{19}$ GeV), naively expected to be at the level of the ratio of the weak and Planck scales ($M_W/M_P\simeq10^{-17}$) or below. The coefficients driving Lorentz invariance violation have been constrained at the $10^{-19}$ level while CPT-violating ones have been limited up to the $10^{-17}$ level. These values, obtained with a reactor-based experiment, are of a similar order as the limits obtained 
by the neutrino-beam experiments LSND~\cite{LSND_LV}, MiniBooNE~\cite{MiniBooNE_LV}, and MINOS (near detector)~\cite{MINOS_LV}. Nevertheless, we emphasize that
the coefficients considered in this analysis are independent and their possible effects have now been studied for the first time. This work completes the coverage of operators in the minimal SME producing neutrino-antineutrino mixing.

\section*{Acknowledgments}
The work of JSD was supported in part by the Department of Energy
under grant DE-FG02-91ER40661 and by the Indiana University Center for Spacetime Symmetries. JMC, JS, and TK are supported by NSF-PHY-1205175. JS is also supported by an MIT Pappalardo Fellowship in Physics.


\begin{thebibliography}{9}
\bibitem{tables}
{\it Data Tables for Lorentz and CPT Violation},
V.A.\ Kosteleck\'y and N.\ Russell,
Rev.\ Mod.\ Phys.\  {\bf 83}, 11 (2011), 2013 edition, arXiv:0801.0287v6.

\bibitem{SBS_LV}
V.A.\ Kosteleck\'y and S. Samuel,
Phys.\ Rev.\ D {\bf 39}, 683 (1989).

\bibitem{LVmodels}
V.A.\ Kosteleck\'y and M.\ Mewes,
Phys.\ Rev.\ D {\bf 70}, 031902(R) (2004);
T.\ Katori {\it et al.}, 
Phys.\ Rev.\ D {\bf 74}, 105009 (2006);
V.\ Barger {\it et al.}, Phys.\ Lett.\ B {\bf 653}, 267 (2007);
J.S.\ D\'\i az and V.A.\ Kosteleck\'y, 
Phys.\ Lett.\ B {\bf 700}, 25 (2011);
Phys.\ Rev.\ D {\bf 85}, 016013 (2012);
V.\ Barger {\it et al.}, Phys.\ Rev.\ D {\bf 84}, 056014 (2011).

\bibitem{KM_SB}
V.A.\ Kosteleck\'y and M.\ Mewes,
Phys.\ Rev.\ D {\bf 70}, 076002 (2004).

\bibitem{DKM}
J.S.\ D\'iaz {\it et al.}, 
Phys. Rev. D {\bf 80}, 076007 (2009).

\bibitem{DC2012}
Y. Abe {\it et al.} [Double Chooz Collaboration], 
Phys. Rev. Lett. {\bf 108}, 131801 (2012);
Phys. Rev. D {\bf 86}, 052008 (2012).

\bibitem{DB}
F.P.\ An {\it et al.} [Daya Bay Collaboration],
Phys. Rev. Lett. {\bf 108}, 171803 (2012).

\bibitem{RENO}
J.K.\ Ahn {\it et al.} [RENO Collaboration],
Phys. Rev. Lett. {\bf 108}, 191802 (2012).

\bibitem{Thiago}
  T.J.C. Bezerra, H.~Furuta, F.~Suekane and T.~Matsubara, Phys. Lett. B {\bf 725}, 271 (2013).

\bibitem{PDG2012}
J. Beringer {\it et al.} 
(Particle Data Group), 
Phys.\ Rev.\ D {\bf 86}, 010001 (2012).

\bibitem{t2k_nim}
  K.~Abe {\it et al.}  [T2K Collaboration],
Nucl. Instrum. Meth. A {\bf 659}, 106 (2011);
Phys. Rev. D {\bf 88}, 032002 (2013).

\bibitem{minos_nue}
P. Adamson {\it et al.} [MINOS Collaboration], 
Phys. Rev. Lett.  {\bf 110}, 171801 (2013).

\bibitem{LVnu}
V.A.\ Kosteleck\'y and M.\ Mewes,
Phys.\ Rev.\ D {\bf 69}, 016005 (2004);
Phys.\ Rev.\ D {\bf 85}, 096005 (2012);
J.S. D\'iaz {\it et al.}, Phys. Rev. D {\bf 88}, 071902 (2013).

\bibitem{DC_LV1}
Y.\ Abe {\it et al.} [Double Chooz Collaboration],
Phys.\ Rev.\ D {\bf 86}, 112009 (2012).

\bibitem{SME}
D.\ Colladay and V.A.\ Kosteleck\'y,
Phys.\ Rev.\ D {\bf 55}, 6760 (1997);
Phys.\ Rev.\ D {\bf 58}, 116002 (1998);
V.A.\ Kosteleck\'y,
Phys.\ Rev.\ D {\bf 69}, 105009 (2004).

\bibitem{IceCube_LV}
R.\ Abbasi {\it et al.}  [IceCube Collaboration],
Phys.\ Rev.\ D {\bf 82}, 112003 (2010).

\bibitem{LSND_LV}
L.B.\ Auerbach {\it et al.} [LSND Collaboration],
Phys.\ Rev.\ D {\bf 72}, 076004 (2005).

\bibitem{MiniBooNE_LV} 
A.A.~Aguilar-Arevalo {\it et al.} [MiniBooNE Collaboration],
Phys.\ Lett.\ B {\bf 718}, 1303 (2013);
T.\ Katori [MiniBooNE Collaboration],
Mod.\ Phys.\ Lett.\ A {\bf 27}, 1230024 (2012).

\bibitem{MINOS_LV}
P.\ Adamson {\it et al.} [MINOS Collaboration],
Phys.\ Rev.\ Lett.\ {\bf 101}, 151601 (2008);
Phys.\ Rev.\ Lett.\ {\bf 105}, 151601 (2010);
Phys.\ Rev.\ D {\bf 85}, 031101 (2012).

\bibitem{RebelMufson} 
B.\ Rebel and S.\ Mufson,
Astropart. Phys. {\bf 48}, 78 (2013).

\bibitem{SunFrame} 
R.~Bluhm {\it et al.},
Phys.\ Rev.\ Lett.\  {\bf 88}, 090801 (2002);
Phys.\ Rev.\ D {\bf 68}, 125008 (2003).

\bibitem{CPTv}
V.A.\ Kosteleck\'y and R.\ Potting,
Nucl. Phys. B {\bf 359}, 545 (1991);
O.W.\ Greenberg,
Phys.\ Rev.\ Lett.\ {\bf 89}, 231602 (2002);
arXiv:1105.0927 [hep-ph].

\bibitem{data_release}
\href{http://doublechooz.in2p3.fr/Scientific/Data_release}{http://doublechooz.in2p3.fr/Scientific/Data\_release}


\bibitem{DC_hydrogen}
Y.~Abe {\it et al.}, 
Phys.~Lett.~B {\bf 723}, 66 (2013).


\bibitem{DC_off}
Y.~Abe {\it et al.}, 
Phys.~Rev.~D {\bf 87}, 011102(R) (2013).


\bibitem{MINUIT}
F. James {\it et al.},
Comput. Phys. Commun. {\bf 10}, 343 (1975).     



\end{thebibliography}
\end{document}